\documentstyle[12pt,epsfig]{article}

\def\be{\begin{equation}}
\def\ee{\end{equation}}
\def\ba{\begin{eqnarray}}
\def\ea{\end{eqnarray}}
\def\ga{\mathrel{\raise.3ex\hbox{$>$\kern-.75em\lower1ex\hbox{$\sim$}}}}
\def\la{\mathrel{\raise.3ex\hbox{$<$\kern-.75em\lower1ex\hbox{$\sim$}}}}

\newcommand{\fr}[2]{\frac{#1}{#2}}

\begin{document}

\baselineskip=16pt
\begin{titlepage}
\rightline{October 2007} \rightline{arXiv:0710.2395}
\begin{center}

\vspace{0.5cm}

\large {\bf Stability of Palatini-f(R) cosmology} \vspace*{5mm}
\normalsize

{\bf Seokcheon Lee$^{\,1}$}

\smallskip
\medskip

$^1${\it Institute of Physics,\\ Academia Sinica,
 Taipei, Taiwan 11529, R.O.C.}

\smallskip
\end{center}

\vskip0.6in

\centerline{\large\bf Abstract}

The evolution of linear cosmological perturbations in modified
theories of gravity is investigated assuming the Palatini
formalism. It has been discussed about the stability problem in
this model based on the equivalence between f(R) gravity and the
scalar tensor theory. However, we study this problem in the
physical frame where the matter is minimally coupled. In general,
the stability of the superhorizon metric evolution depends on
models. We show that the deviation from the superhorizon metric
evolution is null for a specific choice for the nonlinear
Einstein-Hilbert action, $f(\hat{R}) \sim \hat{R}^{n}$, where $n
\neq 0,2,3$. Thus the stability of metric fluctuation is
guaranteed in these models. We also study the matter density
fluctuation in the general gauge and show the differential
equations in super and sub-horizon scales.

\vspace*{2mm}

\end{titlepage}

\section{Introduction}
\setcounter{equation}{0}

The discovery of the present accelerated expansion of the Universe
\cite{SNe, WMAP} can be accounted for either by the existence of a
homogeneous component of energy with a negative pressure, dubbed
dark energy \cite{DE}, or by a modification of gravitational
action with a general function of the scalar curvature instead of
the standard Einstein-Hilbert term, named f(R) gravity \cite{Buc}.

Alternative field equations depend not only on the choice of
action but also on the variational principles \cite{Ferraris}. The
Palatini formalism where the metric and the connections are
treated as independent variables and the energy momentum tensor
does not depend on the independent connection leads to a different
theory from what is obtained from the metric formalism. The
Palatini formalism of f(R) gravity results in second order
differential equations due to the algebraic relation between the
curvature scalar and the trace of the energy momentum tensor.

Cosmological background solutions have been studied for various
gravitational Lagrangian in the Palatini formalism
\cite{background}. Their validity as cosmological theories have
been tested for observation \cite{observ}. Also the cosmological
perturbations have been investigated \cite{perturb}.

It has been known that the Palatini formalism to f(R) gravity is
stable for the curvature scalar perturbation \cite{stability}. The
Newtonian limit of models using the Palatini variational principle
gives contradicting results \cite{Newtonian}. However, the
approach to this problem is using the equivalence between f(R)
gravity and scalar-tensor theory. We will use the evolution of
linear perturbations in f(R) models in the physical frame as
already done in the metric formalism \cite{Hu}.

We study the evolution of linear perturbations in Palatini f(R)
gravity in the physical frame where the matter is minimally
coupled. We consider the stability of metric fluctuations at high
curvature to see the agreement with high redshift cosmological
observations. We also investigate the evolution of the matter
density fluctuation.

In the next section we review the Palatini f(R) gravity. We
investigate the linear perturbation of f(R) gravity in the section
III. Compared with previous works \cite{perturb}, we do not
specify the gauge to find the matter density fluctuation. In
section IV, we derive the stability equation of metric
fluctuations in the high curvature limit and show the stability in
a specific model. We show the evolutions of the metric fluctuation
and the density contrast in the superhorizon and the subhorizon
scales in section V. We reach our conclusions in section VI. We
also provide the detail calculation in the appendix.

\section{Palatini f(R) gravity }
\setcounter{equation}{0}

We consider a modification to the Einstein-Hilbert action assuming
the Palatini formalism, where the metric $g_{\mu\nu}$ and the
torsionless connection $\hat{\Gamma}^{\alpha}_{\mu\nu}$ are
independent quantities and the matter action depends only on
metric \be S = \int d^4 x \sqrt{-g} \Biggl[ \fr{1}{2 \kappa^2} f
\Bigl(\hat{R}(g_{\mu\nu}, \hat{\Gamma}^{\alpha}_{\mu\nu}) \Bigr) +
{\cal L}_{m}(g_{\mu\nu}, \psi) \Biggr] \, , \label{action} \ee
where $\psi$ are matter fields. Then the Ricci tensor is defined
solely by the connection \be \hat{R}_{\mu\nu} =
\hat{\Gamma}^{\alpha}_{\mu\nu , \,\alpha} -
\hat{\Gamma}^{\alpha}_{\mu\alpha  , \,\nu} +
\hat{\Gamma}^{\alpha}_{\alpha\beta}\hat{\Gamma}^{\beta}_{\mu\nu} -
\hat{\Gamma}^{\alpha}_{\mu\beta}\hat{\Gamma}^{\beta}_{\alpha\nu}
\, , \label{hatRmunu} \ee whereas the scalar curvature is given by
\be \hat{R} = g^{\mu\nu} \hat{R}_{\mu\nu} \, . \label{hatR} \ee We
can derive the field equation of f(R) gravity in the Palatini
formalism from the above action (\ref{action}) \be F(\hat{R})
\hat{R}_{\mu\nu} - \fr{1}{2} g_{\mu\nu} f(\hat{R}) = \kappa^2
T_{\mu\nu} \, , \label{seq} \ee where $F(\hat{R}) = \partial
f(\hat{R}) / \partial \hat{R}$ and the matter energy momentum
tensor is given as usual form \be T_{\mu\nu} = - \fr{2}{\sqrt{-g}}
\fr{\delta(\sqrt{-g}{\cal L}_m)}{\delta g^{\mu\nu}} \, .
\label{Tmunu} \ee From the above equations we can get the
generalized Einstein equation \be \hat{G}_{\mu\nu} = G_{\mu\nu} +
\fr{3}{2} \fr{1}{F^2} \nabla_{\mu} F \nabla_{\nu} F - \fr{1}{F}
\nabla_{\mu} \nabla_{\nu} F + g_{\mu\nu} \fr{1}{F} \Box F -
\fr{3}{4} g_{\mu\nu} \fr{1}{F^2} (\partial F)^2 \, ,
\label{hatGmunu} \ee where $\hat{G}_{\mu\nu} \equiv
\hat{R}_{\mu\nu} - \fr{1}{2} g_{\mu\nu} \hat{R}$ and $G_{\mu\nu}
\equiv R_{\mu\nu} - \fr{1}{2} g_{\mu\nu} R$. For the later use, we
reexpress the above equation again as \be F G_{\mu\nu} = \kappa^2
T_{\mu\nu} - \fr{3}{2} \fr{1}{F} \nabla_{\mu} F \nabla_{\nu} F +
\nabla_{\mu} \nabla_{\nu} F + \fr{1}{2} (f - FR) g_{\mu\nu} +
\fr{1}{2} g_{\mu\nu} \Box F \, . \label{FGmunu} \ee From the above
equation we can derive $00$-component of the modified Einstein
equation \be -3 H^2 - 3 HH' = \fr{\kappa^2}{F} \rho + \fr{3}{2}
H^2 \Biggl[ \fr{F''}{F} + \Bigl( 1 + \fr{H'}{H} \Bigr) \fr{F'}{F}
- \fr{F'^2}{F^2} \Biggr] - \fr{1}{2} \fr{f}{F} \, , \label{G001}
\ee where primes denote derivatives with respect to $\ln a$. We
also have the following useful equation \be -2 \fr{H'}{H} =
\fr{\kappa^2}{F H^2} (\rho + p) + \Biggl[ \fr{F''}{F} + \Bigl( -1
+ \fr{H'}{H} \Bigr) \fr{F'}{F} - \fr{3}{2} \fr{F'^2}{F^2} \Biggr]
\, . \label{G002} \ee

\section{Linear perturbation in f(R) gravity (In Conformal Newtonian Gauge)}
\setcounter{equation}{0}

First we start from the metric in the conformal Newtonian gauge.
The line element is given by

\be ds^2 = a^2(\tau) \Biggl[ -\Bigl(1 + 2 \Psi(\tau, \vec{x})
\Bigr) d\tau^2 + \Bigl(1 - 2 \Phi(\tau,\vec{x}) \Bigr) dx^i dx_i
\Biggr] \, . \label{CNG} \ee The main modifications for viable
models with stable high curvature limits happen well during the
matter dominated epoch and we can take the components of the
energy momentum tensor as \be T^{0}_{0} = -\rho (1 + \delta) \, ,
\hspace{0.2in} T^{0}_{i} = \rho \partial_i q \, , \hspace{0.2in}
T^{i}_{j} = 0 \, . \label{Tmunumatter} \ee From the previous
equation (\ref{FGmunu}), we can find the perturbed Einstein
equation \ba F \delta G^{\mu}_{\nu} &=& \kappa^2 \delta
T^{\mu}_{\nu} - R^{\mu}_{\nu} \delta F - \fr{3}{2}
\fr{\delta(\nabla^{\mu} F \nabla_{\nu} F)}{F} + \fr{3}{2}
\fr{\nabla^{\mu} F \nabla_{\nu} F}{F^2} \delta F +
\delta(\nabla^{\mu} \nabla_{\nu} F) \nonumber
\\ && + \Biggl( \fr{3}{2} \fr{\Box F}{F} \delta F + \fr{3}{4}
\fr{\delta (\partial F)^2}{F} - \fr{3}{2} \fr{(\partial F)^2}{F^2}
\delta F - \delta \Box F \Biggr) \delta^{\mu}_{\nu} \, ,
\label{deltaGmunu} \ea where we use the equation (\ref{deltahatR})
and $\delta f(\hat{R}) = F(\hat{R}) \delta \hat{R}$. If we
consider the $ij$-component of the perturbed equation, then we can
find \be \Phi - \Psi = \fr{\delta F}{F} \, , \label{deltaGij} \ee
where we assume the null anisotropic stress. If we use the above
equation (\ref{deltaGij}), then we can express the other
components of the perturbed Einstein equation (\ref{deltaGmunu})
\ba && 3H^2 \Biggl[ \Phi' + \Psi' + \fr{1}{2} \fr{F'}{F} (\Phi' +
\Psi') + \Bigl( \fr{1}{2} \fr{F''}{F} - \fr{1}{2} \fr{F'^2}{F^2} +
\fr{1}{2} \fr{H'}{H} \fr{F'}{F}  + \fr{1}{2} \fr{F'}{F} +
\fr{H'}{H} + 1 \Bigr) \Phi \nonumber \\ && + \Bigl( - \fr{1}{2}
\fr{F''}{F} + \fr{F'^2}{F^2} - \fr{1}{2} \fr{H'}{H} \fr{F'}{F} +
\fr{3}{2} \fr{F'}{F} - \fr{H'}{H} + 1 \Bigr) \Psi
 \Biggr] + \fr{k^2}{a^2} (\Phi + \Psi) =
 - \fr{\kappa^2 \rho}{F} \delta \, , \label{deltaG00c1}
\\ && H \Biggl[\Phi' + \Psi' + \Phi + \Psi + \fr{1}{2} \fr{F'}{F}
(\Phi + \Psi) \Biggr] = - \fr{\kappa^2 \rho}{F} q \, ,
\label{deltaG0ic1} \\ && 3 H^2 \Biggl[ \Phi'' + \Psi'' + \Bigl( 4
+ \fr{H'}{H} \Bigr) \Phi' + \Bigl(3 \fr{F'}{F} + 4 + \fr{H'}{H}
\Bigr) \Psi' + \Bigl(- \fr{F''}{F} - (2 + \fr{H'}{H}) \fr{F'}{F}
\nonumber \\ && + \fr{H'}{H} + 3 \Bigr) \Phi + \Bigl( 3
\fr{F''}{F} + (6 + 3 \fr{H'}{H}) \fr{F'}{F} + 3 \fr{H'}{H} + 3
\Bigr) \Psi \Biggr] = 0 \, . \label{deltaGiic1} \ea To capture the
metric evolution, let us introduce two parameters as in the
reference \cite{Hu}: $\theta$ the deviation from $\zeta$
conservation and $\epsilon$ the deviation from the superhorizon
metric evolution \ba \zeta' = \Phi' + \Psi - H' q &=& - \fr{H'}{H}
\Biggl(\fr{k}{aH} \Biggr)^2 B \theta \, , \label{theta} \\ \Phi''
+ \Psi' - \fr{H''}{H'} \Phi' + \Biggl(\fr{H'}{H} - \fr{H''}{H'}
\Biggr) \Psi &=& - \Biggl(\fr{k}{aH} \Biggr)^2 B \epsilon  \, ,
\label{epsilon} \ea where we define the dimensionless quantity \be
B = \fr{F'}{F} \fr{H}{H'} \, . \label{B} \ee From the above
equations, we can find the expression for $\theta$ and $\epsilon$,
\ba \fr{H'}{H} \Biggl(\fr{k}{aH} \Biggr)^2 B \theta &=& \fr{1}{2}
\Biggl[ \fr{B'}{B} + \fr{3}{2} \fr{H'}{H} B + \fr{H''}{H'} + 4
\Biggr] (\Phi - \Psi) + \fr{1}{2} \Biggl[ 2 H' + \fr{\kappa^2
\rho}{F H} \Biggr] q \nonumber \\
&=& \fr{1}{2} \Biggl[ \fr{B'}{B} + \fr{3}{2} \fr{H'}{H} B +
\fr{H''}{H'} + 4 \Biggr] (\Phi - \Psi) + \fr{1}{2} \Biggl[
-\fr{H'}{H} B' + \Bigl( \fr{H'}{H} - \fr{H''}{H} \Bigr) B
\nonumber \\ && + \fr{1}{2} \fr{H'^2}{H^2} B^2 \Biggr] H q
\label{theta2} \\
\Biggl(\fr{k}{aH} \Biggr)^2 B
 \epsilon &=& \Biggl[ \fr{H'}{H} B + \fr{H''}{H'}
 + \fr{H'}{H} + 3 \Biggr] \Phi' + \fr{1}{2}
 \Biggl[ -2 \fr{B^2}{B^2} - \Bigl(4 \fr{H''}{H'} + 6 \Bigr) \fr{B'}{B}
 + \fr{1}{2} \fr{H'^2}{H^2} B^2 \nonumber \\ && + \fr{5}{2} \fr{H'}{H} B
 - 6 \fr{H''}{H'} - 2 \fr{H''^2}{H'^2} + 3 \fr{H'}{H} + 3 - 2 \Bigl(
 \fr{H''}{H'} + \fr{H'}{H} + 3 \Bigr) \fr{1}{B} \Biggr] \Phi
 \nonumber \\ && +  \fr{1}{2}
 \Biggl[ 2 \fr{B^2}{B^2} + \Bigl(4 \fr{H''}{H'} + 6 \Bigr) \fr{B'}{B}
 + \fr{1}{2} \fr{H'^2}{H^2} B^2 + 2 \fr{H'}{H} B'
 + \Bigl(2 \fr{H''}{H} + \fr{5}{2} \fr{H'}{H} \Bigr) B \nonumber \\ &&
 + 8 \fr{H''}{H'} + 2 \fr{H''^2}{H'^2}
 - \fr{H'}{H} + 3 + 2 \Bigl( \fr{H''}{H'} + \fr{H'}{H} + 3 \Bigr)
 \fr{1}{B} \Biggr] \Psi
\nonumber \\ &=& - \fr{1}{2}
 \Biggl[ 2 \fr{B'^2}{B^2} + \Bigl( 5 \fr{H''}{H'} + \fr{H'}{H} + 9 \Bigr) \fr{B'}{B}
 + \fr{H'}{H} B' + \fr{H'^2}{H^2} B^2 + \Bigl( \fr{5}{2}
 \fr{H''}{H} + \fr{3}{2} \fr{H'^2}{H^2} \nonumber \\ && + 6 \fr{H'}{H} \Bigr) B +
 \fr{H''}{H} + 3 \fr{H''^2}{H'^2} + 13 \fr{H''}{H'} + \fr{H'}{H} +
 9 + 2 \Bigl( \fr{H''}{H'} + \fr{H'}{H} + 3 \Bigr) \fr{1}{B}
 \Biggr] \nonumber \\ && (\Phi - \Psi)  + \Biggl[ \fr{H'}{H} B'
 + \fr{1}{2} \fr{H'^2}{H^2} B^2 +
 \Bigl( \fr{H''}{H} + \fr{3}{2} \fr{H'}{H} \Bigr) B \Biggr] \Psi - \fr{1}{2}
 \Biggl[ \fr{H'}{H} B + \fr{H''}{H'} \nonumber \\ && + \fr{H'}{H} + 3 \Biggr]
 \fr{\kappa^2 \rho}{F H} q  \label{epsilon2}
 \ea From equation (\ref{theta2}), we can recover the conservation
 of Newtonian gauge when $\Phi = \Psi$ and $F$ is a constant.

 In addition to these equations, we can find very useful equation
 from the structure equation (\ref{seq}). By taking the trace of this
 equation (\ref{seq}) and differentiate with $\ln a$, we have \be
 \hat{R}' = \fr{\kappa^2}{F_{,\hat{R}} \hat{R} - F} T' \, .
 \label{hatRprime} \ee Also by taking the perturbation of the
 equation we have \be \delta F \equiv F_{, \hat{R}} \delta \hat{R}
 = \fr{F_{, \hat{R}}}{F_{,\hat{R}} \hat{R} - F} \kappa^2 \delta T
 \, . \label{deltaseq} \ee From these two equations, we can find
 \be \fr{\delta F}{F} = - \fr{1}{3} \fr{F'}{F} \delta = \Phi -
 \Psi \, . \label{deltaF} \ee From this equation we can find the
 evolution equation of matter density fluctuation \ba && \delta'' + \Biggl( 2
 \fr{B'}{B} + \fr{H'}{H} B + 2 \fr{H''}{H'} - \fr{H'}{H} + 3
 \Biggr) \delta' + \Biggl( \fr{H'}{H} B' - 2 \fr{B'^2}{B^2} -
 \Bigl[ 4 \fr{H''}{H'} + 2 \fr{H'}{H} + 4 \Bigr] \fr{B'}{B}
 \nonumber \\ && +
 \fr{H'^2}{H^2} B^2 + \Bigl[ \fr{H''}{H} - \fr{H'^2}{H^2} + 5
 \fr{H'}{H} \Bigr] B + \Bigl[ - 3 \fr{H''}{H} - 4 \fr{H''}{H'} -
 2 \fr{H''^2}{H'^2} + \fr{H'^2}{H^2} - 2 \fr{H'}{H} + 6 \Bigr]
 \nonumber \\ && - 4
 \Bigl[\fr{H''}{H'} + \fr{H'}{H} + 3 \Bigr] \fr{1}{B} \Biggr)
 \delta = - 3 \Psi' \, . \label{deltadoubleprime} \ea Compared
 with previous works \cite{perturb}, we do not specify the gauge
 of matter density to solve the matter density fluctuation.

\section{Stability of metric fluctuations}
\setcounter{equation}{0}

Unstable metric fluctuations can create order unity effects that
invalidate the background expansion history. We can derive the
evolution equation of the deviation parameter. If we differentiate
the equation (\ref{epsilon2}) and consider the evolution in the
superhorizon scale, then we have \ba && \epsilon'' + \Biggl( 2
\fr{B'}{B} + \fr{H'}{H} B + \fr{H''}{H'} - 3 \fr{H'}{H} - 1
\Biggr) \epsilon' + \Biggl( - 2 \fr{B'^2}{B^2} + 2 \fr{H'}{H} B' -
\Bigl[ 5 \fr{H''}{H'} + 4 \fr{H'}{H} + 9 \Bigr] \fr{B'}{B}
\nonumber \\ && + \fr{1}{2} \fr{H'^2}{H^2} B^2 + \Bigl[ 2
\fr{H''}{H'} - 4 \fr{H'^2}{H^2} + \fr{3}{2} \fr{H'}{H} \Bigr] B +
Q' +  \Bigl[ -2 \fr{H''}{H'} - \fr{H'}{H} + 1 \Bigr] Q + 4
\fr{H'^2}{H^2} + 6 \fr{H'}{H} + 7 \nonumber \\
&& - 4 \fr{Q}{B} \Biggr) \epsilon = \fr{1}{B} F(\Psi, \Phi, Hq) \,
, \label{epsilonevol} \ea where we use equations
(\ref{deltaGiic1}) and (\ref{epsilon}) and $F(\Psi, \Phi)$ is the
source function for the deviation $\epsilon$ and define $Q$ as \be
Q = \fr{H''}{H'} + \fr{H'}{H} + 3 \label{Q} \, . \ee The above
equation is different from the metric formalism \cite{Hu}. The
stability of $\epsilon$ depends on the sign of the coefficient of
the term proportional to $\epsilon$. In the metric formalism
$\epsilon$ is stable as long as $B > 0$. However, the stability is
complicate and need to be checked for each model in the Palatini
formalism.

\subsection{A particular example : $ f(\hat{R}) = \beta \hat{R}^{n}$}

We demonstrate the general consideration of the previous
subsection with a specific choice for the nonlinear Lagrangian,
$f(\hat{R}) = \beta \hat{R}^n$, where $n \neq 0, 2, 3$. The
background is simply described by a constant effective equation of
state in this model. The Hubble parameter scales as $H^2 \sim
a^{-3/n}$. Then it is easy to write it with its derivatives in
terms of $\ln a$ \be \fr{H'}{H} = - \fr{3}{2n} \, , \hspace{0.2in}
\fr{H''}{H} = \Biggl( - \fr{3}{2n} \Biggr)^2 \, . \label{H2} \ee
Here the scalar curvature is $\hat{R} = 3(3 - n) H^2 / (2n)$. From
this fact, we can also find the derivatives of $F$ with respect to
$\ln a$ \be \fr{F'}{F} = \fr{F''}{F'} = \fr{3(1 - n)}{n} \, ,
\hspace{0.2in} \fr{F''}{F} = \fr{F'''}{F'} = \Biggl( \fr{3(1 -
n)}{n} \Biggr)^2 \, . \label{F2} \ee If we use above equations
(\ref{H2}) and (\ref{F2}) into (\ref{epsilon2}), then we find that
the deviation from the superhorizon metric evolution is null,
$\epsilon = 0$.

\section{Metric and matter density evolutions}
\setcounter{equation}{0}

\subsection{Superhorizon evolution}
\setcounter{equation}{0}

We consider the metric evolution in superhorizon sized, $k/(aH)
\ll 1$. In this case, the anisotropy relation of the equation
(\ref{theta2}) becomes \be \Phi - \Psi \simeq \Bigl( B + A \Bigr)
H'q \, , \label{superPhi} \ee where $A$ is given by \be A = -
\fr{B \Bigl( 2 B \fr{H'}{H} + 5 \Bigr)}{\Bigl(\fr{B'}{B} +
\fr{3}{2} \fr{H'}{H} B + \fr{H''}{H'} + 4 \Bigr)} \, . \label{A}
\ee From the above equation (\ref{superPhi}), we can find the
superhorizon evolution equation of $\Phi$ from the equation
(\ref{epsilon}) \be \Phi'' + \Biggl( \fr{B'}{B} + 2 \fr{H'}{H} B +
\fr{H''}{H'} - \fr{H'}{H} + 4 - C \Biggr) \Phi' + \Biggl(
\fr{B'}{B} + \fr{H'}{H} B + \fr{H''}{H'} + 3 - C \Biggr) \Phi
\simeq 0 \, , \label{superPhidoubleprime} \ee where $C$ is defined
as \be C = \fr{1}{B + A + 1} \Biggl[ \fr{B'}{B} + 2 \fr{H'}{H} B +
2 \fr{H''}{H'} - \fr{H'}{H} + 3 \Biggr] \, . \label{C} \ee  If we
use this equation (\ref{superPsiprime}), then we have the
evolution equation of matter density (\ref{deltadoubleprime}) \ba
&& \Biggl( 1 + \fr{B}{B + A} \Biggr) \delta'' + \Biggl( 2
 \fr{B'}{B} + \fr{H'}{H} B + 2 \fr{B'A - BA'}{(B + A)^2}
 - \fr{H'}{H} \fr{B}{B + A} + 2 \fr{H''}{H'} - \fr{H'}{H} + 3
 \Biggr) \delta' \nonumber \\ && + \Biggl( \fr{H'}{H} B'
 - 2 \fr{B'^2}{B^2} -
 \Bigl[ 4 \fr{H''}{H'} + 2 \fr{H'}{H} + 4 \Bigr] \fr{B'}{B}
 + \fr{H'^2}{H^2} B^2 + \Bigl[ \fr{H''}{H} - \fr{H'^2}{H^2} + 5
 \fr{H'}{H} \Bigr] B \nonumber \\ && \fr{B''A - BA''}{(B + A)^2}
 - 2 \fr{(B'A -
BA')}{(B + A)^2} \fr{(B' + A')}{(B + A)}  - \fr{H'}{H} \fr{B'A -
BA'}{(B + A)^2} - \Biggl(\fr{H'}{H} \Biggr)' \fr{B}{B + A}
\nonumber \\ && + \Bigl[ - 3 \fr{H''}{H} - 4 \fr{H''}{H'} -
 2 \fr{H''^2}{H'^2} + \fr{H'^2}{H^2} - 2 \fr{H'}{H} + 6 \Bigr] - 4
 \Bigl[\fr{H''}{H'} + \fr{H'}{H} + 3 \Bigr] \fr{1}{B} \Biggr)
 \delta = 0 \, . \label{superdeltadoubleprime} \ea

\subsection{Superhorizon evolution in a particular example}

Now we can check the evolution equations in the previous
subsection in a particular case, $f(\hat{R}) \sim \hat{R}^{n}$. In
this case, we can simplify the following quantities \be B = 2 (n -
1) \, , \hspace{0.2in} A = - 4 (n - 1) = - 2B \, , \hspace{0.2in}
C = \fr{3}{2n} = - \fr{H'}{H} \, . \label{BAC} \ee From this, we
can also simplify the evolution equations
(\ref{superPhidoubleprime}) and (\ref{superdeltadoubleprime}) \ba
\Phi'' + \fr{9 - 4n}{2n} \Phi' &=& 0 \, , \label{sPdp2} \\ \delta'
&=& 0 \, . \label{sddp2} \ea The evolution equation of the
Newtonian potential has no terms proportional to $\Phi$, thus
$\Phi =$ constant is a solution to the equation. Also the matter
density fluctuation has the same for as general relativity,
$\delta =$ constant.

\subsection{Subhorizon evolution}

For subhorizon scales where $k/aH \gg 1$,we can find the Poisson
equation from the equation (\ref{deltaG00c1}) \be k^2 (\Phi +
\Psi) \simeq - \fr{\kappa^2 a^2 \rho}{F} \delta \, .
\label{Poisson} \ee If we use equations (\ref{deltaF}) and
(\ref{Poisson}), then we have \be 3 \Psi \simeq \Biggl( - 3
\fr{\kappa^2 \rho}{F H^2} \fr{a^2 H^2}{k^2} + \fr{F'}{F} \Biggr)
\fr{\delta}{2} \simeq \fr{F'}{F} \fr{\delta}{2} \, .
\label{subPsi} \ee From this equation we can find \be \Phi \simeq
- \Psi \, . \label{subPhi} \ee We can differentiate
(\ref{Phiprime}) and use the above equation (\ref{subPhi}) to get
\ba && \Phi'' + \Biggl(\fr{B'}{B} + \fr{5}{2} \fr{H'}{H} B +
\fr{H''}{H'} + \fr{H'}{H} + 6 \Biggr) \Phi' - \Biggl(
\fr{B'^2}{B^2} + \Bigl[2
\fr{H''}{H'} - 1 \Bigr] \fr{B'}{B} - 2 \fr{H'}{H} B' \nonumber \\
&& - \fr{9}{4} \fr{H'^2}{H^2} B^2  + \Bigl[ 2 \fr{H''}{H} - 4
\fr{H''}{H'} - \fr{21}{2} \fr{H'}{H} \Bigr] B + \Bigl[
\fr{H''^2}{H'^2} - \fr{H''}{H'} + 10 \fr{H'}{H} + 6 \Bigr]
\nonumber \\ && + \Bigl[2 \fr{H''}{H'} + 2 \fr{H'}{H} + 6 \Bigr]
\fr{1}{B} \Biggr) \Phi \simeq 0 \, . \label{subPhidoubleprime} \ea
If we differentiate the equation (\ref{subPsi}) and put into the
equation (\ref{deltadoubleprime}), then we have \ba && \delta'' +
\Biggl( 2
 \fr{B'}{B} + \fr{3}{2} \fr{H'}{H} B + 2 \fr{H''}{H'} - \fr{H'}{H} + 3
 \Biggr) \delta' + \Biggl( \fr{3}{2} \fr{H'}{H} B' - 2 \fr{B'^2}{B^2} -
 \Bigl[ 4 \fr{H''}{H'} + 2 \fr{H'}{H} + 4 \Bigr] \fr{B'}{B}
 \nonumber \\ && +
 \fr{H'^2}{H^2} B^2 + \Bigl[ \fr{3}{2} \fr{H''}{H} - \fr{3}{2} \fr{H'^2}{H^2}
 + 5 \fr{H'}{H} \Bigr] B + \Bigl[ - 3 \fr{H''}{H} - 4 \fr{H''}{H'} -
 2 \fr{H''^2}{H'^2} + \fr{H'^2}{H^2} - 2 \fr{H'}{H} + 6 \Bigr]
 \nonumber \\ && - 4
 \Bigl[\fr{H''}{H'} + \fr{H'}{H} + 3 \Bigr] \fr{1}{B} \Biggr)
 \delta = 0 \, . \label{subdeltadoubleprime} \ea

\subsection{Subhorizon evolution in a particular example}

We can use the previous relation (\ref{BAC}) into the evolution
equations (\ref{subPhidoubleprime}) and
(\ref{subdeltadoubleprime}) \ba \Phi'' + \fr{3(3 - n)}{2n} \Phi'
+ \fr{3(14n^2 + 19n -36)}{4 n^2} \Phi &=& 0 \label{subPdp} \\
\delta'' + \fr{3(2 - n)}{2n} \delta' &=& 0 \label{subddp} \ea Even
though the superhorizon scale evolutions of $\Phi$ and $\delta$
are same to those of general relativity, the subhorizon scale
evolutions of them show different behaviors from those of general
relativity as expected \cite{PZhang}.

\section{Conclusions}

We have analyzed the cosmological evolution of linear
perturbations in Palatini f(R) gravity to see the stability of
metric fluctuations. We have also considered the matter density
fluctuation in the Newtonian gauge.

Compared with metric f(R) gravity, we have shown that the
stability of metric fluctuations in the high redshift limit of
high curvature is not simply expressed. We need to check each
model for the stability. However, we have found that the deviation
from the superhorizon metric evolution is null for a specific
choice of the nonlinear Einstein-Hilbert action, $f(\hat{R}) \sim
\hat{R}^{n}$ and stability of this model is guaranteed.

We have investigated the evolution equations of Newtonian
potential and matter density contrast in super and sub-horizon
scales. In the specific model, superhorizon evolutions of
Newtonian potential and matter density fluctuation are same to
those of general relativity. However, subhorizon evolutions show
the different behaviors from the general relativity case. This
will give us the method to probe the possibility of any
modification of gravity.

\section{Appendix}
\setcounter{equation}{0}
From the equation (\ref{hatRmunu}) we can derive the Ricci tensor
and the scalar curvature by using the metric relation \ba
\hat{R}_{\mu\nu} &=& R_{\mu\nu} + \fr{3}{2} \fr{1}{F^2}
\nabla_{\mu} F \nabla_{\nu} F - \fr{1}{F} \nabla_{\mu}
\nabla_{\nu} F - \fr{1}{2} g_{\mu\nu} \fr{1}{F} \Box F \, ,
\label{hatRmunu2} \\ \hat{R} &=& R - 3 \fr{1}{F} \Box F +
\fr{3}{2} \fr{1}{F^2} (\partial F)^2 \, . \label{hatR2} \ea We can
rewrite field equation (\ref{hatGmunu}) as the form of Einstein
equation plus corrections \be G_{\mu\nu} = \kappa^2 T_{\mu\nu} + (
1 - F) R_{\mu\nu} - \fr{3}{2} \fr{1}{F} \nabla_{\mu} F
\nabla_{\nu} F + \nabla_{\mu} \nabla_{\nu} F + \fr{1}{2} (f - R)
g_{\mu\nu} + \fr{1}{2} g_{\mu\nu} \Box F \, . \label{Gmunu} \ee We
can check that the correction terms are covariantly conserved by
using the useful relation $(\Box \nabla_{\nu} - \nabla_{\nu} \Box)
F = \nabla^{\mu} R_{\mu\nu}$.

Also by taking the trace of the equation (\ref{seq}) we can find
\be 6H^2 + 3 HH' = \fr{\kappa^2}{2F}(- \rho + 3p) - \fr{3}{2} H^2
\Biggl[ \fr{F''}{F} + \Bigl( 3 + \fr{H'}{H} \Bigr) \fr{F'}{F} -
\fr{1}{2} \fr{F'^2}{F^2} \Biggr] + \fr{f}{F} \, . \label{seq2} \ee
By differentiation of the equation (\ref{G002}) we can derive
another useful equation \ba && \fr{F'''}{F} - 3 \fr{F''F'}{F^2} +
\Biggl(3 \fr{H'}{H} + 2 \Biggr) \fr{F''}{F} + \fr{3}{2}
\fr{F'^3}{F^3} - \Biggl( 3 \fr{H'}{H} + \fr{9}{2} \Biggr)
\fr{F'^2}{F^2} + \Biggl( \fr{H''}{H} +
\fr{H'^2}{H^2} + 3 \fr{H'}{H} - 3 \Biggr) \fr{F'}{F} \nonumber \\
&& =  -2 \Biggl( \fr{H''}{H} + \fr{H'^2}{H^2} + 3 \fr{H'}{H}
\Biggr) \label{G003} \ea

The perturbed equations for the Ricci tensor and the scalar
curvature are obtained from the equations (\ref{hatRmunu}) and
(\ref{hatR}) \ba \delta \hat{R}^{\mu}_{\nu} &=& \delta
R^{\mu}_{\nu} + \fr{3}{2 F^2} \delta (\nabla^{\mu} F \nabla_{\nu}
F) - 3\fr{\nabla^{\mu} F \nabla_{\nu} F}{F^3}  \delta F -
\fr{1}{F} \delta( \nabla^{\mu} \nabla_{\nu} F) \nonumber \\ && +
\fr{(\nabla^{\mu} \nabla_{\nu} F)}{F^2} \delta F  + \Biggl( -
\fr{1}{2 F} \delta (\Box F) + \fr{\Box F}{2 F^2} \delta F \Biggr)
\delta^{\mu}_{\nu} \, ,
 \label{deltahatRmunu} \\
\delta \hat{R} &=& \delta R - \fr{3}{F} \delta (\Box F) + 3
\fr{\Box F}{F^2} \delta F + \fr{3}{2 F^2} \delta (\partial F)^2 -
3 \fr{(\partial F)^2}{F^3} \delta F \, . \label{deltahatR} \ea

If we differentiate the equation (\ref{deltaG0ic1}) and use the
equation (\ref{deltaGiic1}), then we have \be \fr{3}{2} \fr{F'}{F}
(\Phi' - \Psi') + \Biggl[ \fr{3}{2} \fr{F''}{F} + \Bigl( \fr{9}{2}
+ \fr{3}{2} \fr{H'}{H} \Bigr) \fr{F'}{F} \Biggr] \Phi +
\Biggl[-\fr{5}{2} \fr{F''}{F} - \Bigl(\fr{7}{2} + \fr{5}{2}
\fr{H'}{H} \Bigr) \fr{F'}{F} - 2 \fr{H'}{H} \Biggr] \Psi =
-\fr{\kappa^2 \rho}{F} \fr{q'}{H} \label{deltaG0ic2} \ee If we use
the equation (\ref{G002}) and adopt $Hq' = - \Psi$, then we can
rewrite the above equation (\ref{deltaG0ic2}) as \be \Phi' - \Psi'
+ \Biggl[ \fr{B'}{B} + \fr{H'}{H} B + \fr{H''}{H'} + 3 \Biggr]
(\Phi - \Psi) - \fr{H'}{H} B \Psi = 0 \, . \label{deltaG0ic3} \ee
We can find $\Phi'$ from the equation (\ref{deltaG0ic1}) and
(\ref{deltaG0ic3}) \be \Phi' = -\fr{1}{2} \Biggl[ \fr{B'}{B} +
\fr{3}{2} \fr{H'}{H} B + \fr{H''}{H'} + 4 \Biggr] (\Phi - \Psi) -
\Psi - \fr{1}{2} \fr{\kappa^2 \rho}{F H} q \, . \label{Phiprime}
\ee

From the equations (\ref{deltaF}) and (\ref{superPhi}) we can
derive \be 3 \Psi = \fr{B}{B + A} \delta' + \Biggl[ \fr{B'A -
BA'}{(B + A)^2} - \fr{H'}{H} \fr{B}{(B + A)} \Biggr] \delta \, .
\label{superPsi} \ee If we differentiate this equation
(\ref{superPsi}), then we have \ba 3 \Psi' &=& \fr{B}{B + A}
\delta'' + \Biggl[ 2 \fr{B'A - BA'}{(B + A)^2} - \fr{H'}{H}
\fr{B}{B + A} \Biggr] \delta' + \Biggl[ \fr{B''A - BA''}{(B +
A)^2} \nonumber \\ && - 2 \fr{(B'A - BA')}{(B + A)^2} \fr{(B' +
A')}{(B + A)}  - \fr{H'}{H} \fr{B'A - BA'}{(B + A)^2} -
\Biggl(\fr{H'}{H} \Biggr)' \fr{B}{B + A} \Biggr] \delta \, .
\label{superPsiprime} \ea


\end{document}